\documentstyle[12pt,amssymb]{article}
\begin{document}
\setlength{\baselineskip}{0.65cm}
\setlength{\parskip}{0.45cm}
\renewcommand{\theequation}{\arabic{equation}}
\renewcommand{\thefootnote}{\fnsymbol{footnote}}
\begin{titlepage}
\normalsize
\vspace{1.5cm}
\begin{center}
\LARGE
{\bf A Non-perturbative Effect in Deep Inelastic Scattering}\\
\vspace{2.5cm}
\large
E.A.\ Paschos\\
\vspace{1.0cm}
Institut f\"ur Theoretische Physik, Universit\"at Dortmund\\
D--44221 Dortmund, Germany\\
and\\
Fermi National Accelerator Laboratory,\\
Batavia, IL 60510, USA
\end{center}
\vspace{3.5cm}
\underline{{\large{Abstract}}}\\ \\
\normalsize
\setlength{\baselineskip}{0.75cm}
\setlength{\parskip}{0.45cm}
The HERA data at large $Q^2$ and small-$x$ investigate
{\underline{large distances on the light-cone}}.  
At such large distances the scattered quarks can 
maintain their colour identity by polarizing the 
vacuum as they transfer energy to it.  We calculate 
the probability for the creation of quark-antiquark 
pairs from the polarized QCD vacuum and their contribution 
to the structure function.
\end{titlepage}
\newpage
The new experiments at HERA \cite{H1,ZEUS} extend the kinematic regions 
of deep
inelastic scattering to much higher values of $Q^2$ and smaller values 
of the Bjorken scattering variable $x$.  We review in this article the
space-time structure of the scattering and show that it involves kinematic
regions where confinement effects are important. 
This motivated me to
consider a new contribution to the structure functions at small $x$.
The analysis considers the rise of $F_2(x)$ at small $x$ and investigates
the additional quanta which are excited.

Deep inelastic scattering studies the tensor
\setcounter{equation}{0}
%
%
\begin{equation}
 W_{\mu\nu}(q\cdot p,q^2)=\frac{1}{2\pi}\int d^4y\,e^{iq\cdot y}
  \langle p|[J_{\mu}(y),\, J_{\nu}(0)]|p\rangle\, .
\end{equation}
The phase of the Fourier transform becomes stationary when
%
%
\begin{equation}
y_- = y_0-y_3 \sim \pm\frac{1}{q_0+q_3} \quad {\rm and} \quad
 y_+ = y_0+y_3 \sim \pm\frac{1}{q_0-q_3}
\end{equation}
which, for the time-like distances investigated by the currents, imply
%
%
\begin{equation}
y^2= y_+y_- \, -y^2_1\, -y^2_2 \leq y^2_0 -y_3^2 \sim \frac{1}{Q^2} \, .
\end{equation}
Thus for large $Q^2$, which is the case at HERA, $y^2$ is very close to
the light cone.  This allows the replacement of the commutator by its
light-cone singularity times a bilocal operator, i.e.,
%
%
\begin{equation}
W_{\mu\nu}(q \cdot p,\, q^2) = s_{\mu\nu\alpha\beta} 
 \int d^4y \, e^{iq \cdot y}
  \frac{\partial}{\partial y^{\alpha}} \, \Delta_F(y) \,
   \langle p|\bar{q}(y)\gamma^{\beta}q(0)|p\rangle\, .
\end{equation}
Now following standard techniques, it can be shown that Bjorken's scaling
follows \cite{Bjorken1}.  This prediction was studied extensively.
Significant violations of scaling have been observed and explained as
perturbative corrections from QCD.  The new
data also indicate that the structure function $F_2(x,Q^2)$ increases by a 
factor of almost two or more as $x$ decreases, signaling the creation of 
additional quanta.
An open issue is still the description of the quanta created at
small $x$.  To this end we note that the distance $y_3$ along the light-cone
becomes very large \cite{Ioffe,Bjorken0}:
%
%
\begin{equation}
y_3 = \pm\frac{1}{2}\, \frac{1}{q_0-q_3} \approx \pm\frac{1}{2M_px} = 2L(x)\, .
\end{equation}
The distance $L(x)$ depends on the scaling variable and for
small $x$ becomes very long in comparison to the size of the proton.
Viewing eq.\ (1) as Compton scattering of two currents on a proton, we are 
forced to accept that the distance between the two currents is many times
the radius of the proton.  Consequently, as the current deposits energy
and momentum on a quark $q'$, it forces it to accelerate and travel a 
long distance ($\sim~\frac{1}{4\,M_p \cdot x} \gg \frac{1}{M_p}$).
In the course of this journey, $q'$ transfers part of its energy to 
the vacuum, e.g.\ by emission of a large number of soft gluons, thus
polarizing it.  Consequently, a chromoelectric flux tube of length
$\sim \frac{1}{4\,M_px}$ is created between the scattered quark $q'$ and 
the rest of the (constituent) spectator quarks, forcing the quark $q'$
to remain confined.  This tube eventually breaks down into two disjoint
tubes when a $q \bar{q}$ pair is created in the field of the tube by the
Schwinger mechanism.  Creation of further pairs creates additional 
breaks, eventually resulting in hadronization.  The number of hadrons
is, roughly, proportional to the number of created $q \bar{q}$ pairs
(multiplicity).  In this article we study the creation of additional 
quark-antiquark pairs from a constant chromoelectric and chromomagnetic 
field \cite{Schwinger1, Schwinger2, Batalin}.  

We represent the states as the product of the scattered quark
plus a number of quark-antiquark pairs in the presence of the QCD vacuum
$|\Omega \rangle$.  The final state, before hadronization, can be written 
schematically as 
%
%
\begin{equation}
\langle X'q'| = \langle \Omega | \left\{ <q'|\, +\, \langle q\bar{q}|<q'|\, 
  + \ldots \, \langle n(q\bar{q})|<q'| \right\}
\end{equation}
where $<q'|$ is the scattered quark, $\langle \Omega|$ the non-perturbative
QCD-vacuum and $\langle n(q\bar{q})|$ a state of $n$-pairs created by the 
Schwinger mechanism. Similarly the initial state is presented as
$|q,\Omega\rangle$. We show
the scattering process in fig.\ 1, where the pairs are created from the 
vacuum.  Energy is transferred to the vacuum through the 
emission of many soft gluons, whose detailed description is not yet available.
Later on we shall substitute the vacuum by a constant chromoelectric and
chromomagnetic field.
\begin{figure}[htb]
\mbox{}
\vskip 2.75in\relax\noindent\hskip -.1in\relax
\includegraphics{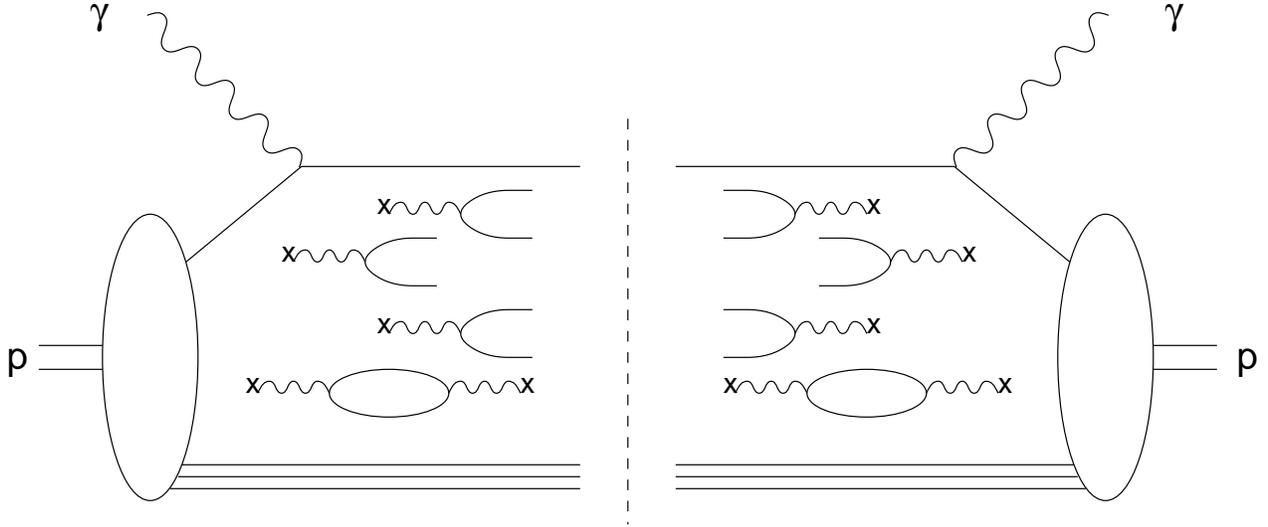} \vskip .4in
\caption{The non-perturbative QCD vacuum with the creation of quark-antiquark
  pairs.}
\end{figure}

I describe next the various terms and processes generated by the S-matrix
of the problem.  The general S-matrix is
%
%
\begin{equation}
S={\rm{\protect{\huge{T}}}}\, exp\,\left\{ -ie \int A_{\mu}(y)\, j^{\mu}(y)\, 
    d^4y -ig\int G^{\alpha}_{\nu}(y)\, j^{\nu ,\alpha}(y)\, d^4y \right\}
\end{equation} 
with $A_{\mu}(y)$ and $G_{\nu}^{\alpha}(y)$ the electromagnetic and
gluonic fields coupled to the corresponding currents 
$j^{\mu}(y) = \bar{q}(y)\gamma^{\mu}q(y)$ and 
$j^{\nu ,\alpha}(y)=\bar{q}(y)\frac{\lambda ^{\alpha}}{2}\gamma ^{\nu}q(x)$,
respectively.  The capital T indicates the time-ordered product.  
We shall treat the electromagnetic interaction perturbatively
and expand its exponential in a series keeping only the term linear in 
$A_{\mu}(y)$.  This can be proven to hold for the time-ordered product.
We write the matrix element as
%
%
\begin{equation}
\langle X',q'|S|q\rangle = \langle X'q'|\,{\rm{T}}\, (-ie)\int A_{\mu}(y) 
  j^{\mu}(y)d^4y\ exp\left\{ -ig \int G_{\nu}^{\alpha}(z)j^{\nu ,\alpha}
    (z)d^4z \right\} |q,\Omega \rangle . 
\end{equation}
Contractions between the quark fields in $j^{\mu}(x)$ and $j^{\mu ,\alpha}(y)$
produce QCD corrections which are frequently included by summing up the
leading logarithmic terms \cite{Fits}.  The perturbative corrections
are characterized by the emission of a few hard gluons from the scattered
quark $q'$, either before or after its collision with the photon.  
Therefore, the quark $q'$, after the emission of these hard gluons and
the collision with the photon, does not create
a long QCD tube.  This part of the quark distribution function,
whose $Q^2$-dependence is determined by the hard-gluon perturbative effects,
will be represented in eq.\ (17) by $q^{pert}_i(x,Q^2)$. The 
distribution by itself is a non-perturbative quantity;
only its $Q^2$-dependence is determined by perturbative QCD.

The new contribution is produced by another class of diagrams generated
also through eq.\ (8), namely those terms which have no contractions 
between the quark fields in $j^{\mu}(y)$ and the quark fields in 
$j^{\nu ,\alpha}(y).$  They involve the contractions of quarks in
$j^{\mu}(y)$ with quarks in the initial and final states.
This permits us to approximate the new matrix element by
%
%
\begin{equation}
\langle X',q'|S|q\rangle _{n-p} \approx
  \langle q'|-ie \int A_{\mu}(y)j^{\mu}(y)d^4y|q\rangle \,\,
   \langle X'|{\rm{T}}\, exp 
    \left\{ -ig\int G^{\alpha}_{\nu}(z)j^{\nu ,\alpha}(z)d^4z \right\}
     |\Omega\rangle
\end{equation}
with the subscript $n-p$ indicating that we treat the strong interaction
term non-\-perturbatively. 
As mentioned already, the quarks in the current $j^{\mu}(y)$ emit many
soft gluons which create the background field (flux-tube).  This field
then is used as the classical field in the QCD vacuum.
The new multiplicative factor from QCD 
involves the transition of the non-perturbative vacuum $|\Omega \rangle$
to any number of final quark pairs.  This transition probability is
related to the amplitude of emitting no pairs, i.e., the vacuum persistence
probability

\begin{displaymath}
S_0 = \langle \Omega |exp \left\{ -ig\int G^{\alpha}_{\nu}(y) j^{\nu ,\alpha}
 (y)d^4y \right\} |\Omega \rangle \, .
\end{displaymath}
We can write
%
%
\begin{equation}
 |S_0|^2 = exp \left\{ -\int d^4y\,\, \omega(y) \right\}
\end{equation}
and identify $\omega(y)$ with the probability for creating a pair per unit 
volume of the flux-tube and per unit time.  
The exact solution of this problem
for a constant electric field was obtained by Schwinger 
\cite{Schwinger1, Schwinger2}.  We adopt
this problem for our flux-tube and consider the potential
%
%
\begin{equation}
G_{\mu}^{\alpha}(y) = (0,\, y_3gB,\, 0,\, tgE)\,\eta^{\alpha}
\end{equation}
to represent the background field with $\eta^{\alpha}$ a unit vector in 
color space. It corresponds to a chromoelectric
field gE along the direction of the tube and a chromomagnetic field gB
perpendicular to the tube.  We found the solution \cite{EAP}
%
%
\begin{equation}
\omega(y,m^2) = \alpha_s \frac{|E|\,\, |B|}{\pi} \sum^{\infty}_{n=1}
  \frac{1}{n}\, {\rm coth} \left( n\frac{|B|}{|E|}\pi \right) \, e^{-
\frac{nm^2\pi}{gE}}
\end{equation}
which in the limit $n\frac{|B|}{|E|}\pi\ll 1$ reduces to the Schwinger
solution \cite{Schwinger1, Schwinger2}.  The non-perturbative 
nature of the solution is manifested
in the exponential function, which has an essential singularity at 
$gE\to 0$.

The picture which emerges so far is a flux-tube of unspecified transverse
dimensions, but whose length is related to the scaling variable through
eq.\ (5).  The quark-antiquark pairs are created in the tube with the
probability density given by eq.\ (12).  The sum of all the pairs modifies
the quark distribution functions by a multiplicative factor.  This will
be a new contribution to be added to the perturbative effects, because,
as explained already, the two terms originate from different contractions
in eq.\ (8). 

We compute next the probability for creating all possible pairs.
To this end we partition the flux tube into small volume elements, as 
shown in fig.\ 2.  The probability of producing a pair in the element 
$dy_i$ and no pair in the rest of the tube is
\begin{figure}[htb]
\mbox{}
\vskip 3.25in\relax\noindent\hskip 1in\relax
\includegraphics{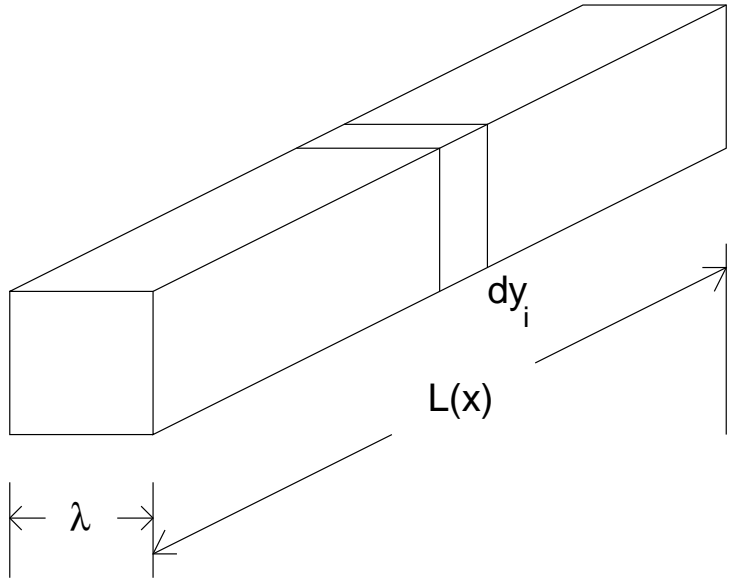} \vskip .4in
\caption{ A schematic drawing of the tube, where pairs and fluctuations are
  created by the background field before and after the interaction with the
  proton.}
\end{figure}
%
%
\begin{equation}
dP_1 = \lambda ^2T\omega (y_i)\, dy_i \prod ^{\infty}_{K=1} 
  \left( 1-\lambda ^2T\omega (y_K)\, dy_K \right)\, .
\end{equation}
We denote the transverse dimension of the tube by $\lambda$ and the time 
for the creation of a pair by $T$.  These are two new parameters to be 
specified later on.  The probability for producing one pair anywhere in the 
tube is
%
%
\begin{equation}
P_1 = \lambda ^2T \int ^{L(x)}_0 \omega (y)dy\,\, e^{-\lambda ^2T\int \omega 
 (y)dy}\, .
\end{equation}
We can generalize this result for $n$-pairs in the tube
%
%
\begin{equation}
P_n(x)=\frac{1}{n!}\left[ \lambda ^2T\int ^{L(x)}_0 \omega (y)dy\right]^n
 e^{-\lambda ^2T \int ^{L(x)}_0 \omega (y)dy} \, .
\end{equation}
Finally, the sum over all possible pairs gives
%
%
\begin{equation}
\sum ^{\infty}_{n=1} P_n(x) = 1-e^{-\lambda ^2T\int ^{L(x)}_0 \omega (y)dy}\, .
\end{equation}
This correction must be multiplied by the distribution function as it follows
from eqs.\ (8) and (9).  Thus each quark distribution function has two 
components:  a perturbative term, which is frequently used to analyze the
data \cite{Fits}, plus a new non-perturbative term 
from the creation of pairs :
%
%
\begin{equation}
q^{total}_i(x) = q^{pert.}_i(x,Q^2) + q^0_i(x_p,Q^2) 
  \left[ 1 - e^{-\lambda ^2T \int^{L(x)}_0 \omega (y)dy} \right]
\end{equation}
The second quark distribution function $q^0_i(x_p,Q^2)$ is generated
by the electromagnetic term in eq.\ (9), where the QCD matrix element
is a multiplicative factor.  This distribution function is not modified
by QCD corrections, a property indicated by the superscript $0$.  For
this reason we expect it to be practically independent of $Q^2$.  
Its numerical value
is determined at an intermediate value of $x=x_p$ where the flux-tube 
begins to form.  Since there are no gluons radiated by the quark we also
expect $q^0_i(x_p,Q^2)$ to remain constant as $x$ decreases.  For the
numerical analysis we select $10^{-2}\leq x_p\leq 10^{-1}$ where the 
structure function is flat in $x$ and independent of $Q^2$.  The additional
factor in the square bracket originates from the creation of pairs.
For large values of $x$ the exponential function is one and this term
vanishes.  For small $x$, where the multiplicity is large, the exponential 
function vanishes and the second term assumes its full strength.
Finally, eq.\ (17) implicitly contains the assumption that the hard gluon
and the tube-like (soft-gluon) effects add up incoherently.  

The distribution of $n$-pairs created from the energy stored
in the vacuum is, according to eq.\ (15), a Poisson distribution.  This 
functional form follows from the property that the creation of pairs in
each cell is independent of what happens in the other cells.  It is also 
independent of
the specific form of $\omega (y)$.  The detailed properties of the pairs
give information beyond the quark distribution functions.\footnote[1]
{The analogous information for the perturbative part, like the 
  probability $P(N)$ of finding a configuration of $N$ partons in the proton 
   or the joint probability of finding partons with longitudinal fractions
    $x_1, \cdots , x_N,$ cannot be calculated as yet \cite{Bjorken2}.}
One consequence is the calculation of the multiplicity
%
%
\begin{equation}
n(x) = \sum ^{\infty}_{n=1}nP_n = \lambda ^2T \int ^{L(x)}_0 \omega (y)dy \, .
\end{equation}
Upon substitution in eq.\ (17) we obtain
%
%
\begin{equation}
q^{total}_i(x,Q^2) = q^{pert.}_i(x,Q^2) + q^0_i(x_p,Q^2) (1-e^{-n(x)}) \, .
\end{equation}
The perturbative term from QCD produces the plateau observed at
$10^{-2} \leq x_p \lesssim 10^{-1}$.  There are several extensions of the
perturbative term to smaller values of $x$.  Among them we must select
one and add on top of it the non-perturbative contribution of the pairs.

We can give an estimate of the effect.  We consider the case
$n\frac{B}{E} \pi \ll 1$ and a constant chromoelectric field.  In this
case the exponent is
%
%
\begin{equation}
n(x) = \lambda ^2T\int ^{L(x)}_0 \omega (y)dy =
 \lambda ^2T \left\{ \frac{\alpha _s E^2}{\pi ^2} \sum ^{\infty}_{n = 1}
  \frac{1}{n^2} e^{\frac{-n\pi m^2}{|gE|}} \right\} L(x) \, . 
\end{equation}
Studies of particle production give the values \cite{Casher}
%
%
\begin{equation}
\frac{1}{2}g_s E = 0.354 GeV^2 \quad {\rm and} \quad 
  \lambda = \sqrt{\pi} \cdot 2.5 \left( \frac{1}{GeV} \right) \, . 
\end{equation}
For these values and $m=m_{\pi}$ the sum in eq.\ (20) is close to 0.80.
We use the uncertainty principle to estimate the lifetime of a virtual
pair as $T \approx \frac{1}{\langle E_{pair}\rangle}$ with 
$\langle E_{pair}\rangle$ the average energy of a pair.  We obtain
%
%
\begin{eqnarray}
n(x) \approx 0.005\, \cdot \frac{\pi}{x}\qquad & {\rm for} & \qquad 
   \langle E_{pair} \rangle = 1.0\,\, GeV \\ \nonumber
{\rm and} \qquad n(x) \simeq 0.003\, \cdot \frac{\pi}{x}\qquad & {\rm for} & 
  \qquad \langle E_{pair} \rangle = 1.5\,\, GeV.
\end{eqnarray}
which implies that pair creation begins to become important for 
$x\sim$ few times $10^{-3}$.  The estimate is very crude, because it 
could be modified by the details of the tube, but still
encouraging because $n(x)$ begins to grow at a value of $x$ close to the
value where the increase of $F_2(x,Q^2)$ is observed.  Alternatively, we
can assume a functional form for $n(x)$ and calculate the structure
function.  A possible functional form is $n(x) = f(x) +\frac{c}{x}$
with $f(x)$ a slowly varying function of $x$ and $c$ a constant.

To sum up, the increase observed in $F_2(x,Q^2)$ may originate 
from the creation
of pairs from the vacuum.  Consequently the increase of $F_2(x,Q^2)$ from
perturbative QCD can be relatively smaller.  As a result a new analysis
of the data is suggested in terms of two components:  a slow increase
from perturbative QCD and a faster increase from the creation of pairs.
The limiting value of $F_2^{non-pert.}(x)$ at $x=0$ is in the present theory finite.
Summing the contributions from all the quarks we obtain
%
%
\begin{equation}
F^{total}_2(x,Q^2) = F^{pert.}_2(x,Q^2) + F_2(x_p,Q^2) (1-e^{-n(x)})
\end{equation}
with $F^{pert.}_2(x,Q^2)$ the perturbative development of the structure 
function and $F_2(x_p,Q^2)$ the structure function measured at
$10^{-2} \leq x_p \leq 10^{-1}$.

The observed increase in $F_2(x,Q^2)$ is closely related to the increase in
the particle multiplicity as $x$ decreases.  The particle multiplicity
as a function of $x$ has not been reported yet.  The data, however, is
available and it is interesting to study the correlation between the 
structure function and the multiplicity expressed in eqs.\ (19) and (23).
The analysis should plot $n(x)\, vs\, x$ and identify a component at
$x\leq 10^{-3}$ which originates from the pairs.  Plotting $n(x)\, vs\, x$
we expect a faster increase setting in a $x \leq 10^{-3}$.

Ideas describing non-vanishing colour fields in the QCD vacuum have
been recognized long time ago and several quantities have already been
studied \cite{Batalin}, \cite{Casher}--\cite{Buch}.  
The novel
aspect of this work is the explanation of the deep inelastic data at
small-x in terms of a chromoelectric tube whose length is determined
by the scaling variable $x$ and the creation of quark-antiquark pairs
by the gluonic field stored in the tube.  The creation of pairs is a
non-perturbative effect which cannot be produced by the exchange of a
finite number of gluons; it comes about as the cumulative effect of
infinite many gluons.\\
\vspace{0.5cm}

\noindent{\Large\bf Acknowledgements} 

\noindent I wish to thank Prof.\ G.\ Savvidy for
his active interest and valuable discussions during the early stages
of this work and Dr.\ G.\ Cvetic for helpful discussions throughout the
work.  I also wish to thank Dr.\ W.\ Bardeen and the theory group at 
Fermilab for their hospitality.  The financial support of the 
Bundesministerium f\"ur Bildung und Forschung under contract No.\
056DO93P(5) is gratefully acknowledged.

\end{document}